\documentclass[twocolumn,english,aps,prd,nofootinbib]{revtex4}
\usepackage{lmodern}
\usepackage[T1]{fontenc}
\usepackage[latin9]{inputenc}
\usepackage{xcolor}
\usepackage{babel}
\usepackage{textcomp}
\usepackage{amsmath}
\usepackage{amsthm}
\usepackage{amssymb}
\usepackage{graphicx}
\usepackage[unicode=true]
 {hyperref}

\makeatletter

\newcommand{\lyxdot}{.}

\@ifundefined{textcolor}{}
{%
 \definecolor{BLACK}{gray}{0}
 \definecolor{WHITE}{gray}{1}
 \definecolor{RED}{rgb}{1,0,0}
 \definecolor{GREEN}{rgb}{0,1,0}
 \definecolor{BLUE}{rgb}{0,0,1}
 \definecolor{CYAN}{cmyk}{1,0,0,0}
 \definecolor{MAGENTA}{cmyk}{0,1,0,0}
 \definecolor{YELLOW}{cmyk}{0,0,1,0}
}
\theoremstyle{plain}
\newtheorem{thm}{\protect\theoremname}
\theoremstyle{remark}
\newtheorem{rem}[thm]{\protect\remarkname}

\usepackage{enumerate}

\makeatother

\providecommand{\remarkname}{Remark}
\providecommand{\theoremname}{Theorem}

\begin{document}
\title{Traversable Morris-Thorne-Buchdahl wormholes in quadratic gravity}
\author{Hoang Ky Nguyen$\,$}
\email[\ \ ]{hoang.nguyen@ubbcluj.ro}

\affiliation{Department of Physics, Babe\c{s}--Bolyai University, Cluj-Napoca
400084, Romania}
\author{Mustapha Azreg-A\"inou$\,$}
\email[\ \ ]{azreg@baskent.edu.tr}

\affiliation{Ba\c{s}kent University, Engineering Faculty, Ba\u{g}lica Campus,
06790-Ankara, Turkey}
\date{July 1, 2023}
\begin{abstract}
\vskip2pt The special Buchdahl-inspired metric obtained in a recent
paper {[}Phys.$\,$Rev.$\,$D \textbf{107}, 104008 (2023){]} describes
asymptotically flat spacetimes in pure $\mathcal{R}^{2}$ gravity.
The metric depends on a new (Buchdahl) parameter $\tilde{k}$ of higher-derivative
characteristic, and recovers the Schwarzschild metric when $\tilde{k}=0$.
It is shown that the special Buchdahl-inspired metric supports a two-way
traversable Morris-Thorne wormhole for $\tilde{k}\in(-1,0)$ in which
case the Weak Energy Condition is formally violated, a naked singularity
for $\tilde{k}\in(-\infty,-1)\cup(0,+\infty)$, and a non-Schwarzschild
structure for $\tilde{k}=-1$.
\end{abstract}
\maketitle

\section{\label{sec:Introduction}Introduction}

Recently, there has been a surge in interest in wormholes, particularly
with the Morris-Thorne ansatz as a guiding principle \citep{MorrisThorne-1988-1,MorrisThorne-1988-2}.
Supermassive objects have been discovered and used as a test bed for
gravity, and although wormholes are of an exotic nature, they may
interact with ordinary matter and can be observed via their astrophysical
signatures \citep{Darmour,Bambi,Azreg,Dzhunushaliev,Cardoso,Konoplya,Nandi,Bueno,Cramer,Nedkova,Harko,Deligianni,Falco,p4,p5,Easson-2015,Easson-2017}.
As a result, it is natural to seek, static and rotating~\citep{p1,p2,p3,p6,p7},
wormholes in modified theories of gravity that permit a violation
of the Weak Energy Condition (WEC) without assuming the existence
of exotic matter.\vskip4pt

All we know about exotic matter is that a) it violates our perception
of energy, that is, an observer may measure some negative amount of
rest energy density and b) it has the ability to sustain wormholes.
So far there has been no general theory about exotic matter and most
of--if not all--the wormhole solutions derived so far were obtained
geometrically upon running the field equations of general relativity
(GR) from left to right: The energy-momentum tensor (EMT) derived
this way was exotic. In this work we rather run the field equations
of purely quadratic gravity in the usual way, that is, by solving
the field equations.\vskip4pt

In generalized and modified theories of GR, explicit EMT are not usually
added to the field equations. Instead, extra terms or corrections
are introduced to the gravitation sector, which can play the role
of exotic matter without truly being exotic matter. An example is
the Brans-Dicke (BD) action, $\int d^{4}x\sqrt{-g}\left[\phi\,\mathcal{R}-\frac{\omega}{\phi}\partial_{\mu}\phi\partial^{\mu}\phi\right]$,
which allows the formation of wormholes \citep{Agnese-1995,Agnese-2001,Campanelli-1993,Vanzo-2012},
where the scalar field $\phi$ acts as an exotic form of matter that
violates the WEC. Another interesting case is the family of $f(\mathcal{R})$
gravity, which in general can be formally cast as a scalar-tensor
theory but the scalar field is directly associated with $\mathcal{R}$
without its own dynamics. As such, the scalar field does not truly
represent an exotic form of matter. In the case of pure $\mathcal{R}^{2}$
gravity, the scalar field is trivially identical to $\mathcal{R}$.
\vskip4pt

Violation of the energy conditions, and particularly of the WEC, has
direct astrophysical consequences. In$\ $\citep{wec1} it was shown
how the non-violation of the WEC constrains the time evolution of
both the Hubble parameter and coordinate distance and sets an upper
bound for $\Omega_{M}$. Similar conclusions were derived in$\ $\citep{wec2}.
In the other case, when the WEC is violated, no such constraints emerge.
Assuming that the SEC and the DEC hold, a first proof of the Cosmic
No-Hair Theorem was given in$\ $\citep{wec3}. Another theorem on
the future fate of a spacetime, the Lorentzian Splitting Theorem,
was proven upon admitting the SEC$\ $\citep{wec4}. From a quantum-mechanical
point of view, all local energy conditions are violated by quantum
fields and also by some classical fields as the non-minimally coupled
scalar fields$\ $\citep{wec5,wec6} and the future-eternal inflating
spacetimes$\ $\citep{wec7}. However, the scale of violation can
be minimized for some cut-and-paste geometric constructions$\ $\citep{wec8},
and for type I wormholes without the cut-and-paste construction$\ $\citep{Azreg}
where the extent of exotic matter has been shown to be inversely proportional
to the square of the mass of the wormhole. A detailed compte-rendu
of the consequences and plausible violations of the energy conditions
are reported in$\ $\citep{wec9}, instances include the possibility
of formation of cosmological singularities in spatially open or flat
spacetimes if the WEC is observed and the possibility of superluminal
motion (wrap drive and traversable wormholes) if the WEC is violated.\vskip4pt

The pure $\mathcal{R}^{2}$ theory was first proposed by Buchdahl
in the early 1960's \citep{Buchdahl-1962} and has recently experienced
a revival of interest, with active investigations into its theoretical
properties as well as its implications for black holes and cosmology
\citep{Lust-2015-backholes,Clifton-2006,Ferreira-2019,Stelle-2015,Pravda-2017,Nguyen-2022-Buchdahl,2023-axisym,2023-WH,Alvarez-2018,Rinaldi-2018,Donoghue-2018,Gurses-2012,Frolov-2009,Stelle-1977,Nguyen-2023-Lambda0,Nguyen-2023-Extension,Nguyen-2023-Nontrivial,Murk-2022,Nelson-2010}.
Among the various extensions of GR, the pure $\mathcal{R}^{2}$ theory
enjoys several distinct advantages. It is a parsimonious theory, having
only one term in the action, $(2\kappa)^{-1}\int d^{4}x\sqrt{-g}\,\mathcal{R}^{2}$,
where the gravitational coupling $\kappa$ is a dimensionless parameter.
It is the only theory that is both scale invariant and ghost-free.
Regarding the former, it actually possesses a restricted scale invariance
under a Weyl transformation, $g_{\mu\nu}\rightarrow\Omega^{2}(x)g_{\mu\nu}$,
where $\Omega(x)$ obeys a harmonic condition, $\square\,\Omega=0$,
as discovered in \citep{Edery-2014}. Regarding the latter, as a member
of $f(\mathcal{R})$, its scalar degree of freedom involves only second-order
derivatives when transitioning from the Jordan frame to the Einstein
frame, thereby evading the Ostrogradsky instability that often plagues
higher-derivative gravity \citep{Woodard}. Furthermore, pure $\mathcal{R}^{2}$
gravity has been shown propagate two massless modes: a spin-2 tensor
mode and a spin-0 scalar mode, each carrying it own significance \citep{AlvarezGaume-2015}.
On one hand, the massless spin-2 tensor mode indicates the emergence
of a long-range potential with the correct Newtonian tail $\sim1/r$
\citep{Nguyen-2023-Newtonian}. On the other hand, the massless spin-0
scalar mode could potentially be responsible for an additional long-range
potential, thereby introducing new physics.\vskip4pt

One concrete realization of new physics in pure $\mathcal{R}^{2}$
gravity manifests through the work of Buchdahl in 1962, in which he
originated a program aimed at finding vacuum solutions for the theory
\citep{Buchdahl-1962}. He was able to make significant progress with
his efforts boiling down to solving a non-linear second-order ordinary
differential equation (ODE). If an analytical solution to his ODE
could be found, then the vacuo solutions he sought would automatically
ensue. Unfortunately, Buchdahl deemed the ODE insoluble, prompting
him to suspend further pursuit. Consequently, his groundbreaking paper
has remained relatively obscure within the gravitational research
community for the past sixty years. However, recent advancements made
by one of us have revitalized Buchdahl's program and brought it to
fruition. Section \ref{sec:Buchdahl-inspired-spacetimes} in this
paper will review its final outcome.\vskip4pt

Significantly, the Buchdahl-inspired solutions exhibit non-constant
scalar curvature, a distinctive feature resulting from the fourth-derivative
nature of the theory. This non-constant scalar curvature is controlled
by a new parameter known as the Buchdahl parameter $k$. Remarkably,
these solutions defy the generalized Lichnerowicz ``theorem'' proposed
in \citep{Nelson-2010,Stelle-2015,Lust-2015-backholes}, which stipulates
that static vacuum solutions of pure $\mathcal{R}^{2}$ gravity must
possess constant scalar curvature exclusively. The Buchdahl-inspired
solutions evade this ``theorem'' by circumventing one of the central
assumptions \citep{Nguyen-2023-Extension}. The non-constant scalar
curvature observed in these solutions is a manifestation of higher-derivative
effects, which are encapsulated by the Buchdahl parameter $k$.\textcolor{blue}{{}
\vskip4pt}

In this paper, we shall use the closed analytical vacuum solution
for pure $\mathcal{R}^{2}$ gravity derived in Ref.$\ $\citep{Nguyen-2023-Lambda0}
to show the formation of a wormhole that connects two asymptotically
flat spacetime sheets via a ``throat''. This wormhole is enabled
by the high-derivative nature of the theory without requiring complicated
ingredients or true exotic matter.\vskip4pt

The paper is structured as follows. In Sec. \ref{sec:Buchdahl-inspired-spacetimes}
we review Buchdahl-inspired metrics obtained in Refs.$\ $\citep{Nguyen-2022-Buchdahl,Nguyen-2023-Lambda0,Nguyen-2023-Nontrivial};
in Sec. \ref{sec:Two-new-representations} we present two additional
representations for Buchdahl-inspired metrics that are asymptotically
flat; in Sec. \ref{sec:MTB-metric} we map the \emph{special} (asymptotically
flat) Buchdahl-inspired metrics to the Morris-Thorne ansatz, investigate
their properties, and construct a wormhole when the Weak Energy Condition
is violated.

\section{\label{sec:Buchdahl-inspired-spacetimes}Buchdahl-inspired \large{$\mathcal R^2$} \small{spacetimes}}

In Ref.$\ $\citep{Nguyen-2022-Buchdahl} we advanced a program initiated
by Buchdahl in 1962 seeking vacuo configurations for pure $\mathcal{R}^{2}$
gravity \citep{Buchdahl-1962}. The field equation in vacuo 
\begin{equation}
\mathcal{R}\left(\mathcal{R}_{\mu\nu}-\frac{1}{4}g_{\mu\nu}\mathcal{R}\right)+\left(g_{\mu\nu}\square-\nabla_{\mu}\nabla_{\nu}\right)\mathcal{R}=0
\end{equation}
has a static spherisymmetric solution which is expressible in terms
of two auxiliary functions $p(r)$ and $q(r)$ per 
\begin{equation}
ds^{2}=e^{k\int\frac{dr}{r\,q(r)}}\left\{ -\frac{p(r)q(r)}{r}dt^{2}+\frac{p(r)\,r}{q(r)}dr^{2}+r^{2}d\Omega^{2}\right\} 
\end{equation}
The two functions $p$ and $q$ are coupled via a pair of first-order
evolution-type ODE's: 
\begin{align}
\frac{dp}{dr} & =\frac{3\,k^{2}}{4\,r}\frac{p}{q^{2}}\\
\frac{dq}{dr} & =(1-\Lambda\,r^{2})\,p
\end{align}
Reflecting the fourth-order nature of quadratic gravity, this solution
is specified by \emph{four} parameters: $\Lambda$, $k$, $p_{0}$
and $q_{0}$. When $k=0$ the evolution rules recover the Schwarzschild-de
Sitter. When $k\neq0$, the Ricci scalar is \emph{non-constant} and
is given by 
\begin{equation}
\mathcal{R}(r)=4\Lambda\,e^{-k\int\frac{dr}{r\,q(r)}}
\end{equation}
It approaches $4\Lambda$ at spatial infinity, indicating an asymptotic
de Sitter behavior.\vskip4pt

In Ref.$\ $\citep{Nguyen-2023-Lambda0} we further advanced the solutions
for the case $\Lambda=0$ and obtained an exact closed analytical
form for an asymptotically flat non-Schwarzschild metric, which was
called the \emph{special} Buchdahl-inspired metric, expressible as
\footnote{Note that we used a different set of notations for variables in that
paper.} 
\begin{align}
ds^{2} & =\left|1-\frac{r_{\text{s}}}{r}\right|^{\tilde{k}}\times\nonumber \\
 & \ \ \biggl\{-\left(1-\frac{r_{\text{s}}}{r}\right)dt^{2}+\frac{dr^{2}}{1-\frac{r_{\text{s}}}{r}}\frac{\rho^{4}(r)}{r^{4}}+\rho^{2}(r)d\Omega^{2}\biggr\}\label{eq:special-B-1}
\end{align}
\begin{equation}
\rho^{2}(r):=\zeta^{2}r_{\text{s}}^{2}\,\frac{\left|1-\frac{r_{\text{s}}}{r}\right|^{\zeta-1}}{\left(1-\text{sgn}\left(1-\frac{r_{\text{s}}}{r}\right)\left|1-\frac{r_{\text{s}}}{r}\right|^{\zeta}\right)^{2}}\label{eq:special-B-2}
\end{equation}
in which $\tilde{k}:=\frac{k}{r_{\text{s}}}$ and $\zeta:=\sqrt{1+3\tilde{k}^{2}}$.
It contains two parameters, $r_{\text{s}}$ playing the role of a
Schwarzschild radius, and $\tilde{k}$ a new (Buchdahl) dimensionless
parameter. The solution holds for all value of $r\in\mathbb{R}$ except
at $r=0$ and $r=r_{\text{s}}$. The radial direction thus comprises
of three sections:\vskip4pt

1. The ``exterior'', $r>r_{\text{s}}$,\vskip4pt

2. The ``interior'', $0<r<r_{\text{s}}$,\vskip4pt

3. The ``repulsive'' gravity domain, $r<0$. We exclude this unphysical
region from our consideration.\vskip8pt

Note that the two components $g_{tt}$ and $g_{rr}$ flip their signs
at the interior-exterior boundary, $r=r_{\text{s}}$. The Kruskal-Szekeres
diagram is analytically constructed in Ref.$\ $\citep{Nguyen-2023-Lambda0}.\vskip4pt

Although the special Buchdahl-inspired metric, Eqs. \eqref{eq:special-B-1}
and \eqref{eq:special-B-2}, is Ricci-scalar flat, viz. $\mathcal{R}=0$,
it is not Ricci flat, hence non-Schwarzschild. Moreover, it can be
verified that \citep{Nguyen-2023-Nontrivial}
\begin{equation}
\mathcal{R}^{-1}\nabla_{\mu}\nabla_{\nu}\mathcal{R}=\mathcal{R}_{\mu\nu}\neq0
\end{equation}
and (upon taking the trace) 
\begin{equation}
\mathcal{R}^{-1}\,\square\,\mathcal{R}=\mathcal{R}=0
\end{equation}
That is to say, the solution formally obeys the following equation
\begin{equation}
G_{\mu\nu}:=\mathcal{R}_{\mu\nu}-\frac{1}{2}g_{\mu\nu}\mathcal{R}=\mathcal{R}^{-1}\nabla_{\mu}\nabla_{\nu}\mathcal{R}\label{eq:non-trivial}
\end{equation}
with the \emph{non-vanishing} term in the right hand side acting as
a ``quasi'' energy-momentum tensor (EMT) and making the solution
non-Schwarzschild. This ``quasi'' EMT is thus a surrogate of exotic
matter which would sustain a wormhole under certain circumstances
to be explored in this paper.

\section{\label{sec:Two-new-representations}Two new representations for asymptotically
flat Buchdahl-inspired metrics}

\subsection{The isotropic coordinates}

For the ``exterior'' section, let us choose a variable $\bar{r}$
and a function $g(\bar{r})$ to fulfill two requirements: 
\begin{align}
\rho^{2}(r) & =g(\bar{r})\bar{r}^{2},\\
\frac{\rho^{4}(r)}{\left(1-\frac{r_{\text{s}}}{r}\right)r^{4}}\left(\frac{dr}{d\bar{r}}\right)^{2} & =g(\bar{r})
\end{align}
With $\rho(r)$ given in Eq. \eqref{eq:special-B-2}, solving them
\footnote{
\begin{equation}
\int dx\frac{x^{a-1}}{1-x^{2a}}=\frac{1}{a}\int\frac{d(x^{a})}{1-x^{2a}}=\frac{1}{2a}\ln\frac{1+x^{a}}{1-x^{a}}
\end{equation}
} 
\begin{equation}
\frac{d\bar{r}}{\bar{r}}=\frac{\rho(r)}{\left(1-\frac{r_{\text{s}}}{r}\right)^{\frac{1}{2}}}\frac{dr}{r^{2}}=\zeta\,\frac{\left(1-\frac{r_{\text{s}}}{r}\right)^{\frac{\zeta-2}{2}}}{1-\left(1-\frac{r_{\text{s}}}{r}\right)^{\zeta}}d\left(1-\frac{r_{\text{s}}}{r}\right)
\end{equation}
giving 
\begin{equation}
\bar{r}=r_{*}\frac{1+\left(1-\frac{r_{\text{s}}}{r}\right)^{\frac{\zeta}{2}}}{1-\left(1-\frac{r_{\text{s}}}{r}\right)^{\frac{\zeta}{2}}}
\end{equation}
or 
\begin{equation}
r=r_{\text{s}}\left(1-\left|\frac{1-\frac{r_{*}}{\bar{r}}}{1+\frac{r_{*}}{\bar{r}}}\right|^{\frac{2}{\zeta}}\right)^{-1}
\end{equation}
and 
\begin{align}
g(\bar{r}) & =\frac{\rho^{2}(r)}{\bar{r}^{2}}=\frac{\zeta^{2}r_{\text{s}}^{2}}{\bar{r}^{2}}\,\frac{\left(1-\frac{r_{\text{s}}}{r}\right)^{\zeta-1}}{\left(1-\left(1-\frac{r_{\text{s}}}{r}\right)^{\zeta}\right)^{2}}\\
 & =\frac{\zeta^{2}r_{\text{s}}^{2}}{16r_{*}^{2}}\left(1+\frac{r_{*}}{\bar{r}}\right)^{4}\left|\frac{1-\frac{r_{*}}{\bar{r}}}{1+\frac{r_{*}}{\bar{r}}}\right|^{\frac{2}{\zeta}(\zeta-1)}\\
 & =\left(1-\frac{\zeta^{2}r_{\text{s}}^{2}}{16\bar{r}^{2}}\right)^{2}\left|\frac{1-\frac{\zeta r_{\text{s}}}{4\bar{r}}}{1+\frac{\zeta r_{\text{s}}}{4\bar{r}}}\right|^{-\frac{2}{\zeta}}\ \ \ \text{choosing }r_{*}=\frac{\zeta\,r_{\text{s}}}{4}
\end{align}
rendering the metric

{\small{}
\begin{align}
ds^{2} & =-\left|\frac{1-\frac{\zeta r_{\text{s}}}{4\bar{r}}}{1+\frac{\zeta r_{\text{s}}}{4\bar{r}}}\right|^{\frac{2}{\zeta}(\tilde{k}+1)}dt^{2}\nonumber \\
 & \ \ \ \,+\left|\frac{1-\frac{\zeta r_{\text{s}}}{4\bar{r}}}{1+\frac{\zeta r_{\text{s}}}{4\bar{r}}}\right|^{\frac{2}{\zeta}(\tilde{k}-1)}\left(1-\frac{\zeta^{2}r_{\text{s}}^{2}}{16\bar{r}^{2}}\right)^{2}\left(d\bar{r}^{2}+\bar{r}^{2}d\Omega^{2}\right)
\end{align}
}{\small\par}

It is straightforward to see that these expressions are unchanged
upon the ``image reflection'' 
\begin{equation}
\frac{4\bar{r}}{\zeta r_{\text{s}}}\leftrightarrows\frac{\zeta r_{\text{s}}}{4\bar{r}}
\end{equation}
This means that the two \emph{separate} domains $\bar{r}>\frac{\zeta r_{\text{s}}}{4}$
and $\bar{r}<\frac{\zeta r_{\text{s}}}{4}$ are reciprocal images,
with the value $\zeta r_{\text{s}}/4$ being a ``reflection point''.
Also, the case $\tilde{k}=0$ (viz. $\zeta=1$) recovers the Schwarzschild
metric in Weyl's isotropic coordinates: 
\begin{equation}
ds^{2}=-\left(\frac{1-\frac{r_{\text{s}}}{4\bar{r}}}{1+\frac{r_{\text{s}}}{4\bar{r}}}\right)^{2}dt^{2}+\left(1+\frac{r_{\text{s}}}{4\bar{r}}\right)^{4}\left(d\bar{r}^{2}+\bar{r}^{2}d\Omega^{2}\right)\label{eq:iso-Schwa}
\end{equation}

\subsection*{Kretschmann invariant}

The Kretschmann scalar $K:=\mathcal{R}^{\mu\nu\rho\sigma}\mathcal{R}_{\mu\nu\rho\sigma}$
is given in \citep{Nguyen-2023-Lambda0} and we shall not reproduce
it here. We only report its expression for the isotropic coordinates
which, by design, only cover the ``exterior'' section\vskip-4pt

{\small{}
\begin{align}
K & =\frac{2}{\zeta^{8}r_{\text{s}}^{4}}\left|\frac{1+\frac{\zeta r_{\text{s}}}{4\bar{r}}}{1-\frac{\zeta r_{\text{s}}}{4\bar{r}}}\right|^{4\left(2+\frac{\tilde{k}-1}{\zeta}\right)}\left(1-\left(\frac{1-\frac{\zeta r_{\text{s}}}{4\bar{r}}}{1+\frac{\zeta r_{\text{s}}}{4\bar{r}}}\right)^{2}\right)^{6}\times\nonumber \\
 & \Biggl\{4\tilde{k}^{2}(\tilde{k}+1)\left(\frac{1-\frac{\zeta r_{\text{s}}}{4\bar{r}}}{1+\frac{\zeta r_{\text{s}}}{4\bar{r}}}\right)^{2}\nonumber \\
 & +\zeta\,\Bigl(4\tilde{k}^{3}-5\tilde{k}^{2}-3\Bigr)\left(1-\left(\frac{1-\frac{\zeta r_{\text{s}}}{4\bar{r}}}{1+\frac{\zeta r_{\text{s}}}{4\bar{r}}}\right)^{4}\right)\nonumber \\
 & +\Bigl(9\tilde{k}^{4}-2\tilde{k}^{3}+10\tilde{k}^{2}+3\Bigr)\left(1+\left(\frac{1-\frac{\zeta r_{\text{s}}}{4\bar{r}}}{1+\frac{\zeta r_{\text{s}}}{4\bar{r}}}\right)^{4}\right)\Biggr\}\label{eq:a-19c}
\end{align}
}{\small\par}
\begin{rem}
Interestingly, the ``non-analytic'' piece in $K$ is isolated in
$\left|\frac{1+\frac{\zeta r_{\text{s}}}{4\bar{r}}}{1-\frac{\zeta r_{\text{s}}}{4\bar{r}}}\right|^{4\left(2+\frac{\tilde{k}-1}{\zeta}\right)}$
. Since $2\zeta+\tilde{k}-1>0\ \ \forall\tilde{k}\in\mathbb{R}$,
this non-analytic piece is solely responsible for the divergence of
$K$ at the reflection point $\frac{\zeta\,r_{\text{s}}}{4}$ (except
for $\tilde{k}=0$ and $\tilde{k}=-1$, see below). 
\end{rem}
\vskip6pt 
\begin{rem}
Only for $\tilde{k}=0$ $(\zeta=1$) and $\tilde{k}=-1$ $(\zeta=2)$,
does the exponent $\frac{4}{\zeta}(2\zeta+\tilde{k}-1)$ equal $4$.
The non-analytic piece gets canceled by the terms $\left(\frac{1-\frac{\zeta r_{\text{s}}}{4\bar{r}}}{1+\frac{\zeta r_{\text{s}}}{4\bar{r}}}\right)^{4}$
insides the curly bracket. For $\tilde{k}=0$ $(\zeta=1)$ 
\begin{equation}
K=12r_{\text{s}}^{2}\frac{\bar{r}^{6}}{(\bar{r}+r_{\text{s}}/4)^{12}}
\end{equation}
whereas for $\tilde{k}=-1$ $(\zeta=2)$ 
\begin{equation}
K=\frac{3}{8}r_{\text{s}}^{2}\frac{\bar{r}^{6}}{(\bar{r}+r_{\text{s}}/2)^{12}}
\end{equation}
In both cases, the Kretschmann scalar carries the same function form. 
\end{rem}

\subsection{Another representation}

Let us define a new radial coordinate $r'$ such that 
\begin{equation}
1-\frac{\zeta r_{\text{s}}}{r'}:=\text{sgn}\left(1-\frac{r_{\text{s}}}{r}\right)\left|1-\frac{r_{\text{s}}}{r}\right|^{\zeta}
\end{equation}
Then 
\begin{align}
\frac{dr'}{r'^{2}} & =\left|1-\frac{r_{\text{s}}}{r}\right|^{\zeta-1}\frac{dr}{r^{2}}
\end{align}
and 
\begin{equation}
\rho^{2}(r)=r'^{2}\left|1-\frac{\zeta r_{\text{s}}}{r'}\right|^{\frac{\zeta-1}{\zeta}}
\end{equation}
The metric given in \eqref{eq:special-B-1}--\eqref{eq:special-B-2}
can be brought into 
\begin{align}
ds^{2} & =-\text{sgn}\left(1-\frac{\zeta r_{\text{s}}}{r'}\right)\left|1-\frac{\zeta r_{\text{s}}}{r'}\right|^{\frac{\tilde{k}+1}{\zeta}}dt^{2}\nonumber \\
 & \ \ \ +\text{sgn}\left(1-\frac{\zeta r_{\text{s}}}{r'}\right)\left|1-\frac{\zeta r_{\text{s}}}{r'}\right|^{\frac{\tilde{k}-1}{\zeta}}dr'^{2}\nonumber \\
 & \ \ \ +\left|1-\frac{\zeta r_{\text{s}}}{r'}\right|^{\frac{\tilde{k}-1}{\zeta}+1}r'^{2}d\Omega^{2}
\end{align}
This representation brings the special Buchdahl-inspired metric under
the umbrella of the \emph{generalized} Campanelli-Lousto solution
in Brans-Dicke gravity that we uncover in another report \citep{2023-WEC}.

\section{\label{sec:MTB-metric}Morris-Thorne-Buchdahl wormholes}

\begin{figure*}[!t]
\noindent \begin{centering}
\includegraphics[scale=0.99]{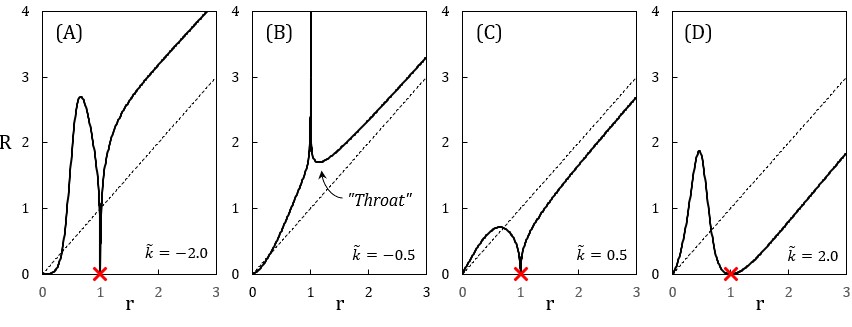}
\par\end{centering}
\caption{\label{fig:R(r)-special}$R$ vs $r$ for the \emph{special} Buchdahl-inspired
metric; $r_{\text{s}}=1$. Panel (B), representative of $\tilde{k}\in(-1,0)$,
yields a minimum for $R(r)$ and corresponds to a wormhole.}
\end{figure*}

In terms of $x:=1-\frac{r_{\text{s}}}{r}\in\mathbb{R}$, the special
Buchdahl-inspired metric becomes {\small{}
\begin{align}
ds^{2} & =-\text{sgn}(x)\left|x\right|^{\tilde{k}+1}dt^{2}+\zeta^{4}r_{\text{s}}^{2}\frac{\text{sgn}(x)\left|x\right|^{\tilde{k}+2\zeta-3}}{(1-\text{sgn}(x)\left|x\right|^{\zeta})^{4}}dx^{2}\nonumber \\
 & \ \ \ \ \ \ \ \ +\zeta^{2}r_{\text{s}}^{2}\frac{\left|x\right|^{\tilde{k}+\zeta-1}}{(1-\text{sgn}(x)\left|x\right|^{\zeta})^{2}}d\Omega^{2}\label{eq:MT-ansatz-in-x}
\end{align}
}{\small\par}

The areal radius in the class is 
\begin{equation}
R=\zeta r_{\text{s}}\frac{\left|x\right|^{\frac{1}{2}(\tilde{k}+\zeta-1)}}{1-\text{sgn}(x)\left|x\right|^{\zeta}}\label{eq:areal-R-in-x}
\end{equation}

For the exterior: {\small{}
\begin{align}
\frac{dR}{dr} & =\frac{\zeta r_{\text{s}}^{2}}{2r^{2}}\frac{x^{\frac{1}{2}(\tilde{k}+\zeta-3)}}{\left(1-x^{\zeta}\right)^{2}}\left[(\tilde{k}-1+\zeta)-(\tilde{k}-1-\zeta)x^{\zeta}\right]
\end{align}
}which would have an acceptable root{\small{}
\begin{equation}
x_{\text{ext}}=\left(\frac{\tilde{k}-1+\sqrt{1+3\tilde{k}^{2}}}{\tilde{k}-1-\sqrt{1+3\tilde{k}^{2}}}\right)^{\frac{1}{\sqrt{1+3\tilde{k}^{2}}}}\in(0,1)\label{eq:xext}
\end{equation}
}if $\tilde{k}\in(-1,0)$.\vskip4pt

For the interior: {\small{}
\begin{align}
\frac{dR}{dr} & =-\frac{\zeta r_{\text{s}}^{2}}{2r^{2}}\frac{(-x)^{\frac{1}{2}(\tilde{k}+\zeta-3)}}{\left(1+(-x)^{\zeta}\right)^{2}}\left[(\tilde{k}-1+\zeta)+(\tilde{k}-1-\zeta)(-x)^{\zeta}\right]
\end{align}
}which would have an acceptable root{\small{}
\begin{equation}
x_{\text{int}}=-\left(-\frac{\tilde{k}-1+\sqrt{1+3\tilde{k}^{2}}}{\tilde{k}-1-\sqrt{1+3\tilde{k}^{2}}}\right)^{\frac{1}{\sqrt{1+3\tilde{k}^{2}}}}\in(-\infty,0)
\end{equation}
}if $\tilde{k}\in(-\infty,1)\cup(0,+\infty)$.\vskip4pt

The behavior of $R$ as a function of $r$ is shown in \ref{fig:R(r)-special}.
Panel (B) is representative of $\tilde{k}\in(-1,0)$ exhibits a minimum
for $R(r)$ in the exterior.\vskip4pt

We shall bring the metric above to the Morris-Thorne ansatz \citep{MorrisThorne-1988-2}
\begin{equation}
ds^{2}=-e^{2\Phi(R)}dt^{2}+\frac{dR^{2}}{1-\frac{b(R)}{R}}+R^{2}d\Omega^{2}\label{eq:MT-metric}
\end{equation}
If we focus on the ``exterior'' region alone, then $(x\in(0,1))$
\begin{equation}
ds^{2}=-x^{\tilde{k}+1}dt^{2}+\zeta^{4}r_{\text{s}}^{2}\frac{x^{\tilde{k}+2\zeta-3}}{(1-x^{\zeta})^{4}}dx^{2}+\zeta^{2}r_{\text{s}}^{2}\frac{x^{\tilde{k}+\zeta-1}}{(1-x^{\zeta})^{2}}d\Omega^{2}
\end{equation}
For the exterior, viz. $x\in(0,1)$, let us make a further coordinate
transformation 
\begin{equation}
y:=x^{\zeta}\in(0,1)
\end{equation}
and denoting $A:=\frac{\tilde{k}+1}{\zeta}$ and $B:=\frac{\tilde{k}-1}{\zeta}$
(again, $\zeta=\sqrt{1+3\tilde{k}^{2}}$) 
\begin{align}
ds^{2} & =-y^{A}dt^{2}+\left(\zeta\,r_{\text{s}}\right)^{2}\frac{y^{B}}{(1-y)^{2}}\left[\frac{dy^{2}}{(1-y)^{2}}+y\,d\Omega^{2}\right]
\end{align}
In summary, the areal radius is defined as 
\begin{equation}
R=\zeta\,r_{\text{s}}\frac{y^{\frac{B+1}{2}}}{1-y}\label{eq:areal-radius}
\end{equation}
The redshift function and the shape function are given by, respectively,
\begin{align}
e^{2\Phi(R)} & =y^{A},\label{eq:redshift-func}\\
1-\frac{b(R)}{R} & =\frac{(1-y)^{4}}{\zeta^{2}r_{\text{s}}^{2}y^{B}}\left(\frac{dR}{dy}\right)^{2}\\
 & =\frac{1}{4y}\left[(B-1)y-(B+1)\right]^{2}\label{eq:shape-func}
\end{align}
for the region $y\geqslant y_{*}$ where 
\begin{equation}
y_{*}=\frac{B+1}{B-1}
\end{equation}
corresponds to $x_{ext}$ in Eq. \eqref{eq:xext}.

\subsection*{The four Morris-Thorne constraints}

In the exterior, $r>r_{\text{s}}$, hence $x\in(0,1)$, taking derivative
of Eq. \eqref{eq:areal-R-in-x}: 
\begin{equation}
\frac{dR}{dy}=\frac{\zeta\,r_{\text{s}}y^{\frac{B-1}{2}}}{2(1-y)^{4}}\left[(B+1)-(B-1)y\right]
\end{equation}
The equation $\frac{dR}{dx}=0$ has a single root at 
\begin{equation}
y_{*}=\frac{B+1}{B-1}\label{eq:y-star-def}
\end{equation}
This root is acceptable, viz. $y_{*}\in(0,1)$ if 
\begin{equation}
B<-1\ \ \ \text{or}\ \ \ -1<\tilde{k}<0
\end{equation}
The minimum 
\begin{equation}
R_{*}=\frac{\zeta\,r_{\text{s}}}{2}(1-B)^{\frac{1-B}{2}}(-1-B)^{\frac{1+B}{2}}\label{eq:R-star}
\end{equation}

\emph{Constraint \#1}.---The redshift function $\Phi(R)$ (given
in \eqref{eq:redshift-func}) be finite everywhere (hence no horizon).\vskip6pt

\emph{Constraint \#2.}---Minimum value of the $R$-coordinate, i.e.
at the throat of the wormhole, $R_{*}$ being the minimum value of
$R$, given in Eq. \eqref{eq:R-star}.\vskip6pt

\emph{Constraint \#3.}---Finiteness of the proper radial distance,
i.e. $b(R)/R\leq1$ (for $R\geq R_{*}$) throughout the space. The
equality sign holds only at the throat. This is required in order
to ensure the finiteness of the proper radial distance $l(R)$ given
in \eqref{eq:proper-l} where the \textpm{} signs refer to the two
asymptotically flat regions which are connected by the wormhole. Note
that the condition $b(R)/R\le1$ assures that the metric component
$g_{RR}$ does not change its sign for any $R\ge R_{*}$.\vskip6pt

\emph{Constraint \#4.}---Asymptotic flatness condition, i.e. as $l\rightarrow\pm\infty$
(or equivalently, $R\rightarrow\infty$ or $r\rightarrow\infty$ or
$x\rightarrow1^{-}$) then $b(R)/R\rightarrow0$.

\subsection*{Embedding}

\begin{figure}[!t]
\begin{centering}
\includegraphics[scale=0.64]{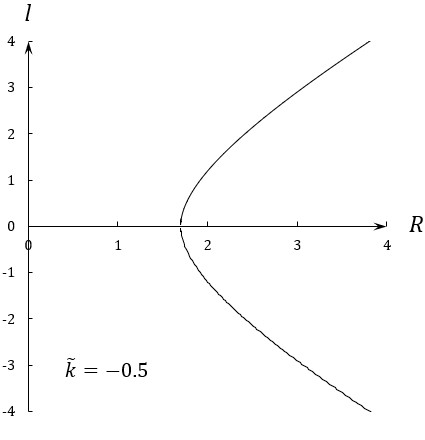}
\par\end{centering}
\begin{centering}
\includegraphics[scale=0.5]{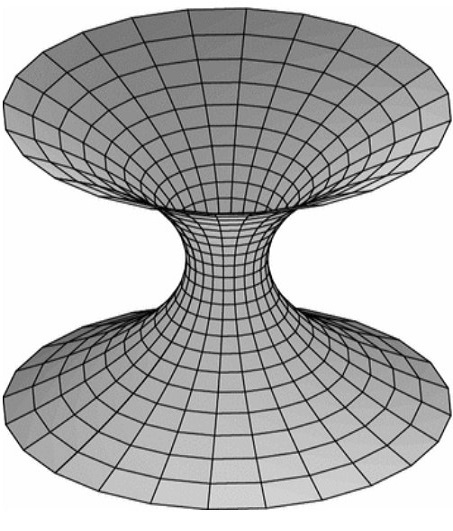}
\par\end{centering}
\caption{\label{fig:Proper-distance}Proper radial distance (upper panel) and
embedding diagram (lower panel).}
\end{figure}
\begin{figure*}[!t]
\noindent \begin{centering}
\includegraphics[scale=0.5]{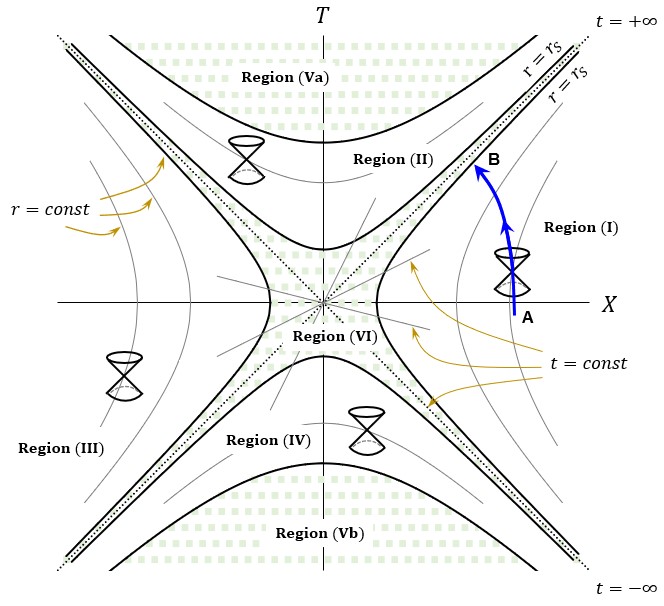}$\ \ \ \ \ $\includegraphics[scale=0.5]{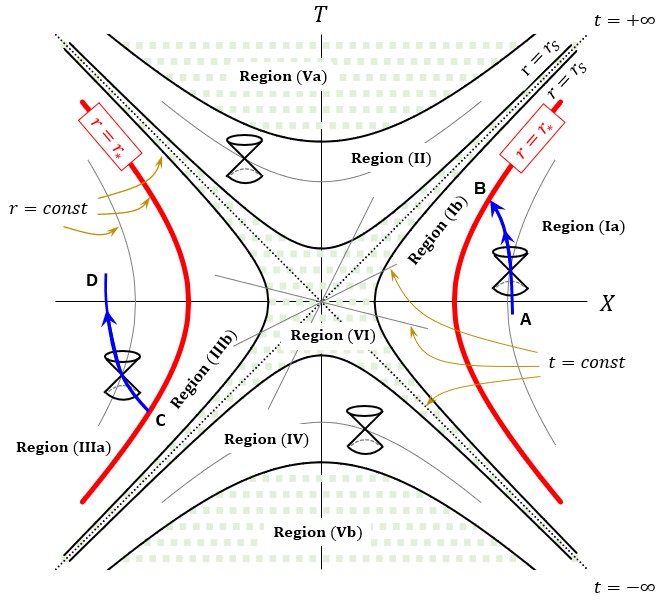}$\ \ \ \ \ $
\par\end{centering}
\caption{\label{fig:KS-diagram} $\ \zeta-$Kruskal-Szekeres diagrams for the
asymptotically flat Buchdahl-inspired spacetimes for $\tilde{k}\in(-\infty,-1)\cup(0,+\infty)$
(the case of naked singularity, shown in left panel) and for $\tilde{k}\in(-1,0)$
(the case of wormhole, shown in right panel). Left panel: On the infalling
radial timelike trajectory (blue line), a particle in Region (I) eventually
hits the naked singularity, $r=r_{\text{s}}$. $\ $Right panel: The
wormhole ``throat'' is depicted by the red lines, which further split
Region (I) into (Ia) and (Ib), and Region (III) into (IIIa) and (IIIb).
On the radial trajectory $A\rightarrow B\rightarrow C\rightarrow D$,
a particle in Region (Ia) first enters the wormhole mouth at point
B (hence, an infalling motion) then escape into Region (IIIa) by emerging
at the other mouth at point C (hence, on outgoing motion). As the
two red lines are \textquotedblleft glued\textquotedblright{} together
to form a wormhole that connects Region (Ia) and Region (IIIa), the
two opposite points B and C represent the same spacetime event.}
\end{figure*}

With $y_{*}$ defined in Eq. \eqref{eq:y-star-def}, the shape function
in Eq. \eqref{eq:shape-func} is 
\begin{equation}
1-\frac{b(R)}{R}=\frac{(B-1)^{2}}{4y}\left(y-y_{*}\right)^{2}\geq0
\end{equation}

In the embedding diagram, the MT ansatz 
\begin{equation}
ds^{2}=-e^{2\Phi(R)}dt^{2}+\left(1+\left(\frac{dz}{dR}\right)^{2}\right)dR^{2}+R^{2}d\Omega^{2}\label{eq:embed}
\end{equation}
yields 
\begin{align}
\frac{dz}{dR} & =\pm\frac{1}{\sqrt{\frac{R}{b(R)}-1}}=\pm\frac{\sqrt{b(R)/R}}{\sqrt{1-b(R)/R}}\\
 & =\pm\frac{\sqrt{4y-(B-1)(y-y_{*})^{2}}}{(B-1)(y-y_{*})}
\end{align}
Obviously $dz/dR$ diverges at $y=y_{*}$, meaning that in the embedding
diagram (the lower panel of Fig. \ref{fig:Proper-distance}), $z(R)$
is vertical at $y=y_{*},$the ``throat'' of the wormhole. The function
$z(R)$ is a combination of Appell hypergeometric functions that is
not particularly illuminating and hence will not be produced here.
\footnote{Specifically, $\ \ z(R)=\pm\zeta\,r_{\text{s}}\int_{y_{*}}^{y}dy\frac{y^{\frac{B}{2}}}{(1-y)^{2}}\sqrt{1-\frac{((B-1)y-(B+1))^{2}}{4y}}$.}\vskip4pt

Nevertheless, the proper radial distance is simpler to obtain: {\small{}
\begin{align}
l(R) & =\pm\int_{R_{*}}^{R}\frac{dR}{\sqrt{1-\frac{b(R)}{R}}}\\
 & =\pm\zeta\,r_{\text{s}}\int_{y_{*}}^{y}dy\frac{y^{\frac{B}{2}}}{(1-y)^{2}}\\
 & =\pm\frac{\zeta\,r_{\text{s}}}{1+B/2}\times\biggl[y^{1+B/2}\,_{2}F_{1}\Bigl(2,1+B/2;2+B/2;y\Bigr)\nonumber \\
 & \ \ \ \ \ \ \ \ \ \ \ -y_{*}^{1+B/2}\,_{2}F_{1}\Bigl(2,1+B/2;2+B/2;y_{*}\Bigr)\biggr]\label{eq:proper-l}
\end{align}
}As an example, the upper panel of Fig. \ref{fig:Proper-distance}
plots the proper radial distance for $\tilde{k}=-0.5$, viz. $B\approx-1.134$,
$y_{*}\approx0.0628$, $r_{*}\approx1.14\,r_{\text{s}}$, $R_{*}\approx1.7\,r_{\text{s}}$.
A ``throat'' is manifest at $R=R_{*}$. \vskip4pt

In the range of $\tilde{k}\in(-\infty,-1)\cup(0,+\infty)$, The Kretschmann
invariant diverges at $y=0$ (i.e., $r=r_{\text{s}}$), indicating
a physical singularity on the interior-exterior boundary, $y=0$.
As a result, the spacetime is not geodesically complete, and the geodesics
terminate at the physical singularity. The $\zeta-$Kruskal-Szekeres
(KS) diagram previously constructed in Ref.$\ $\citep{Nguyen-2023-Lambda0}
is reproduced in Fig.$\ $\ref{fig:KS-diagram} here for the reader's
convenience. In the left panel of Fig.$\ $\ref{fig:KS-diagram},
we also show the radial infalling motion of a massive particle along
the trajectory $A\rightarrow B$, with the particle eventually hitting
the physical singularity, represented by point $B$ on the interior-exterior
boundary. \vskip4pt

In the range of $\tilde{k}\in(-1,0)$, the Kretschmann invariant likewise
diverges at $y=0$ (i.e., $r=r_{\text{s}}$), indicating a physical
singularity on the interior-exterior boundary, $y=0$. However, since
the areal radius possesses a minimum value at $y_{*}=\frac{B+1}{B-1}\in(0,\,1)$,
we can generate a wormhole solution by ``gluing'' the region $y_{*}\leq y<1$
(corresponding to $r_{*}\leq r<+\infty$ as shown in Panel (B) of
Fig.~\ref{fig:R(r)-special}, with $r_{*}=\frac{r_{\text{s}}}{1-y_{*}^{1/\zeta}}>r_{\text{s}}$)
with its symmetric counterpart (also in the region $y_{*}\leq y<1$)
in the $\zeta-$KS diagram. The right panel of Fig.$\ $\ref{fig:KS-diagram}
shows how this is done. The wormhole ``throat'' is represented by
the two red lines, $r=r_{*}$, which further split Region (I) into
(Ia) and (Ib), and Region (III) into (IIIa) and (IIIb). That is to
say, we are connecting an asymptotically flat exterior sheet (i.e.,
Region (Ia)) with another asymptotically flat exterior sheet (i.e.,
Region (IIIa)), which are mirror images of each another with respect
to a sign flip $(T,X)\leftrightarrow(-T,-X)$ in their $\zeta-$KS
coordinates. \vskip4pt

In this construction, both the upper and lower sheets of the wormhole
correspond to $r_{*}\leq r<+\infty$ and they approach asymptotic
flatness at spatial infinity. The two sheets are smoothly connected
at the ``throat'' at $y=y_{*}$, where $z(R)$ becomes vertical,
hence ensuring a smooth connection. The upper and lower sheets are
distinguished by the $\pm$ sign in the proper length parameter $l$,
per Eq. \eqref{eq:proper-l}, which runs continuously from $-\infty$
to $\infty$ as a traveler moves from the lower sheet of the wormhole
to the upper one. \vskip4pt

In the right panel of Fig.$\ $\ref{fig:KS-diagram}, the radial infalling
motion of a massive particle is depicted by the trajectory $A\rightarrow B$
in Region (Ia), with $B$ lying on the ``throat''. The particle then
emerges at point $C$ (which also lies on the ``throat'' and is opposite
to point $B$ on the $\zeta-$KS diagram) then continue on the path
$C\rightarrow D$ in Region (IIIa). Note that points $B$ and $C$
represent the same spacetime event.\textcolor{blue}{{} \vskip8pt}

\subsection*{Violation of the Weak Energy Condition}

Formally, the geometric form for the Weak Energy Condition requires
that $G_{\mu\nu}t^{\mu}t^{\nu}\geq0$ for every future-pointing timelike
vector $t^{\mu}$; e.g., see Ref.$\ $\citep{Koutou}. In particular,
$G_{00}\geq0$.\vskip4pt

The special Buchdahl-inspired metric has the $00-$component of the
Einstein tensor 
\begin{equation}
G_{00}=\frac{\tilde{k}(\tilde{k}+1)}{2r_{\text{s}}^{2}\zeta^{4}}\frac{\left[1-\text{sgn}\left(1-\frac{r_{\text{s}}}{r}\right)\left|1-\frac{r_{\text{s}}}{r}\right|^{\zeta}\right]^{4}}{\left|1-\frac{r_{\text{s}}}{r}\right|^{2(\zeta-1)}}\label{eq:G00}
\end{equation}
For $\tilde{k}\in(-1,0)$, the exterior region exhibits a wormhole
``throat'' as can be seen in Panel (B) of Fig. \ref{fig:R(r)-special}.
At the same time, $G_{00}<0\ \ \forall r$ for $\tilde{k}\in(-1,0)$,
thus violating the Weak Energy Condition. \vskip4pt

As Morris and Thorne envisioned \citep{MorrisThorne-1988-1}, in a
traversable wormhole, light rays that enter it at one mouth then reemerge
at its other mouth have a cross-sectional area initially decreasing
and then increasing. In order for this phenomenon to occur, there
necessarily be some ``gravitational repulsion'' near the ``throat'',
exerting influence on the light rays. In Eq. \eqref{eq:G00}, the
magnitude of $G_{00}$ monotonically increases as one approaches the
interior-exterior boundary, $r=r_{\text{s}}$. For $\tilde{k}\in(-1,0)$,
the negative and dominant $G_{00}$ component thus acts like gravitational
repulsion. We expect that it could leave signatures, distinguishing
a wormhole from a black hole \citep{distinguish}.\vskip4pt

Simultaneously, the interior-exterior boundary is a set of naked singularities.
Consequently, the spacetime for $\tilde{k}\in(-1,0)$ accommodates
both a wormhole connecting two asymptotically flat exterior sheets
(viz. Regions (Ia) and (IIIa) in Fig.$\ $\ref{fig:KS-diagram}) and
a set of naked singularities (belonging to Regions (Ib) and (IIIb)),
thereby representing a non-trivial geometrical structure in this situation.
\vskip4pt

As mentioned in the concluding remark of Section \ref{sec:Buchdahl-inspired-spacetimes},
the right hand side of Eq.$\ $\eqref{eq:non-trivial} corresponds
to a ``quasi'' energy-momentum tensor, defined as (modulo a multiplicative
constant)
\begin{equation}
T_{\text{\ensuremath{\mu\nu}}}:=\mathcal{R}^{-1}\nabla_{\mu}\nabla_{\nu}\mathcal{R}
\end{equation}
It was shown in Ref.$\ $\citep{Nguyen-2023-Nontrivial} that despite
the vanishing Ricci scalar throughout spacetime, this ``quasi'' energy-momentum
tensor remains well-defined and is identical to the $G_{00}$ component
presented in Eq. \eqref{eq:G00}. It effectively serves as a surrogate
to exotic matter required to sustain a wormhole.

\section{\label{sec:Conclusion}Conclusion}

In a previous work \citep{Nguyen-2023-Lambda0}, we derived a special
Buchdahl-inspired metric that describes asymptotically flat spacetimes
in pure $\mathcal{R}^{2}$ gravity. This metric, expressed in a closed
analytical form, enabled us to construct a Kruskal-Szekeres diagram
representing the maximal analytic extension for the metric. The Buchdahl
parameter $\tilde{k}$ in the metric is a new parameter that reflects
the higher-derivative nature of the pure $\mathcal{R}^{2}$ action.\vskip4pt

In this paper, we present several additional advancements. Firstly,
we describe two additional representations of the metric. Secondly,
we examine the metric within the framework of the Morris-Thorne ansatz.
For values of $\tilde{k}$ falling within the ranges $(-\infty,-1)$
and $(0,+\infty)$, the interior-exterior boundary constitutes a naked
singularity. However, for the range $\tilde{k}\in(-1,0)$, the areal
radius has a minimum value in exterior region. Despite the geodesic
incompleteness of the solution in this situation, where geodesics
terminate on the singularity, it is possible to shield the singularity
by removing the region of space neighboring the singularity and gluing
the Kruskal-Szekeres symmetric copy of the remaining space region
to that same region. The Morris-Thorne-Buchdahl wormhole constructed
this way consists of a pair of asymptotically flat spacetime sheets
connected at a ``throat'' that allows two-way passage.\vskip4pt

Thirdly, we find that when a wormhole is formed, that is, when $\tilde{k}\in(-1,0)$,
the Weak Energy Condition is formally violated, even though no exotic
matter is in presence. Therefore, pure $\mathcal{R}^{2}$ theory can
support a wormhole without the need for truly exotic matter in the
energy-momentum tensor or complicated ingredients in the gravitation
sector such as torsion, non-metricity, or non-locality. Our work opens
up new avenues for exploring the fascinating properties of wormholes
and naked singularities in higher-derivative gravity theories. For
wormholes and naked singularities of quadratic relativity, all known
potential astrophysical observations, including light deflection,
precession, shadow, and quasi-periodic oscillations are the subjects
of a future investigation~\citep{future} and cannot be carried out
in this work.\vskip4pt

Another important question, even more important than all that has
been said, is the stability of the wormhole discussed in this work.
Stability analysis is a more involved issue~\citep{st1,st2,s1,s2,s3}
as this necessitates to perform a perturbation analysis of the metric,
which will make the subject of another subsequent paper. However,
based on the generic analysis made in~\citep{s4} the wormhole is
likely to be stable.
\begin{acknowledgments}
We thank Tiberiu Harko for his helpful suggestions during the development
of this research. We thank the anonymous referee for highly constructive
comments.
\begin{figure}[t]
\noindent \begin{centering}
\includegraphics[scale=0.75]{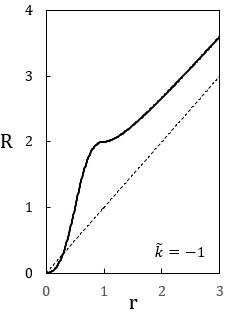}
\par\end{centering}
\caption{\label{fig:R(r)-borderline}$R$ vs $r$ of the \emph{special} Buchdahl-inspired
metric for $\tilde{k}=-1$, $r_{\text{s}}=1$.}
\end{figure}
\end{acknowledgments}

\begin{center}
-----------------$\infty$----------------- 
\par\end{center}

\appendix

\section{$\,$The borderline case of $\tilde{k}=-1$}

For completeness, we shall briefly examine the $\tilde{k}=-1$ case,
viz. $\zeta=2$. With $x:=1-\frac{r_{\text{s}}}{r}$, from \eqref{eq:areal-R-in-x},
the areal radius is 
\begin{equation}
R=\frac{2r_{\text{s}}}{1-\text{sgn}(x)x^{2}}=\frac{2r_{\text{s}}}{1-\text{sgn}\left(1-\frac{r_{\text{s}}}{r}\right)\left(1-\frac{r_{\text{s}}}{r}\right)^{2}}
\end{equation}
which is a monotonic increasing function for $r\in\mathbb{R}^{+}$
and $R(r=r_{\text{s}})=2r_{\text{s}}$, as shown in Fig. \ref{fig:R(r)-borderline}.
In this coordinate, the metric \eqref{eq:MT-ansatz-in-x} becomes
\begin{equation}
ds^{2}=-\text{sgn}\left(1-\frac{2r_{\text{s}}}{R}\right)dt^{2}+\frac{dR^{2}}{1-\frac{2r_{\text{s}}}{R}}+R^{2}\,d\Omega^{2}
\end{equation}
which differs from the Schwarzschild metric by the $g_{tt}$ component.
The equations of motion (EOM) in the exterior, viz. $R>2r_{\text{s}}$,
are 
\begin{equation}
\frac{d}{d\tau}\left(g_{\sigma\lambda}\frac{dx^{\lambda}}{d\tau}\right)=\frac{1}{2}\partial_{\sigma}g_{\mu\nu}\frac{dx^{\mu}}{d\tau}\frac{dx^{\nu}}{d\tau}
\end{equation}
or (with $\mathcal{E}$ and $l$ being two constants of motions, the
dot denoting derivative with respect to $\tau$, and restricting to
the $\theta=\pi/2$ plane)

\begin{align}
\dot{t} & =\mathcal{E}\ \ \ \ \ \text{for }\sigma=\lambda=0\\
\dot{\varphi} & =\frac{l}{R^{2}}\ \ \ \ \ \text{for }\sigma=\lambda=2
\end{align}
subject to a constraint 
\begin{align}
-\dot{t}^{2}+\frac{\dot{R}^{2}}{1-\frac{2r_{\text{s}}}{R}}+R^{2}\dot{\varphi}^{2} & =f=\begin{cases}
\ \ 0 & \text{for null geodesics}\\
-1 & \text{for timelike geodesics}
\end{cases}
\end{align}
We subsequently get 
\begin{align}
\dot{R}^{2} & =\left(1-\frac{2r_{\text{s}}}{R}\right)\left(f+\mathcal{E}^{2}-\frac{l^{2}}{R^{2}}\right)\\
 & =f+\mathcal{E}^{2}-\frac{l^{2}}{R^{2}}-\frac{2r_{\text{s}}(f+\mathcal{E}^{2})}{R}+\frac{2r_{\text{s}}l^{2}}{R^{3}}
\end{align}
and 
\begin{equation}
\ddot{R}=\frac{r_{\text{s}}(f+\mathcal{E}^{2})}{R^{2}}+l^{2}\left(\frac{1}{R^{3}}-\frac{3r_{\text{s}}}{R^{4}}\right)
\end{equation}
For a massive object $(f=-1)$: 
\begin{equation}
\ddot{R}=-(1-\mathcal{E}^{2})\frac{r_{\text{s}}}{R^{2}}+l^{2}\left(\frac{1}{R^{3}}-\frac{3r_{\text{s}}}{R^{4}}\right)
\end{equation}
Compared with the EOM in Schwarzschild $\ddot{R}=-\frac{r_{\text{s}}}{2R^{2}}+l^{2}\left(\frac{1}{R^{3}}-\frac{3r_{\text{s}}}{2R^{4}}\right)$,
the only modification is the Newtonian potential being ``renormalized''
by $(1-\mathcal{E}^{2})$ and thus acting like ``repulsive'' force
if $\mathcal{E}^{2}>1$. It is worth mentioning that this is a partial
result of much wider conclusions drawn in~\citep{future} for any
$\tilde{k}$, among which the velocity-dependent acceleration for
massive objects.


\begin{thebibliography}{10}
\bibitem{MorrisThorne-1988-1}M.$\,$S. Morris and K.$\,$S. Thorne,
\emph{Wormholes in spacetime and their for interstellar travel: A
tool for teaching general relativity}, Am. J. Phys. \textbf{56}, 5
(1988)

\bibitem{MorrisThorne-1988-2}M.$\,$S. Morris, K.$\,$S. Thorne,
and U. Yurtsever, \emph{Wormholes, Time Machines, and the Weak Energy
Condition}, Phys. Rev. Lett. \textbf{61}, 1446 (1988)

\bibitem{Darmour}T. Damour and S.$\,$N. Solodukhin, \emph{Wormholes
as black hole foils}, Phys. Rev. D \textbf{76}, 024016 (2007), \textcolor{purple}{\href{https://arxiv.org/abs/0704.2667}{arXiv:0704.2667 [gr-qc]}}

\bibitem{Bambi}C. Bambi, \emph{Can the supermassive objects at the
centers of galaxies be traversable wormholes? The first test of strong
gravity for mm/sub-mm very long baseline interferometry facilities},
Phys. Rev. D \textbf{87}, 107501 (2013), \textcolor{purple}{\href{https://arxiv.org/abs/1304.5691}{arXiv:1304.5691 [gr-qc]}}

\bibitem{Azreg}M. Azreg-A\"inou, \emph{Confined-exotic-matter wormholes
with no gluing effects - Imaging supermassive wormholes and black
holes}, JCAP \textbf{07}, 037 (2015), \textcolor{purple}{\href{https://arxiv.org/abs/1412.8282}{arXiv:1412.8282 [gr-qc]}}

\bibitem{Dzhunushaliev}V. Dzhunushaliev, V. Folomeev, B. Kleihaus,
and J. Kunz, \emph{Can mixed star-plus-wormhole systems mimic black
holes?}, JCAP \textbf{08}, 030 (2016), \textcolor{purple}{\href{https://arxiv.org/abs/1601.04124}{arXiv:1601.04124 [gr-qc]}}

\bibitem{Cardoso}V. Cardoso, E. Franzin, and P. Pani, \emph{Is the
Gravitational-Wave Ringdown a Probe of the Event Horizon?}, Phys.
Rev. Lett. \textbf{116}, 171101 (2016), \textcolor{purple}{\href{https://arxiv.org/abs/1602.07309}{arXiv:1602.07309 [gr-qc]}}

\bibitem{Konoplya}R.$\,$A. Konoplya and A. Zhidenko, \emph{Wormholes
versus black holes: quasinormal ringing at early and late times},
JCAP \textbf{12}, 043 (2016), \textcolor{purple}{\href{https://arxiv.org/abs/1606.00517}{arXiv:1606.00517 [gr-qc]}}

\bibitem{Nandi}K.$\,$K. Nandi, R.$\,$N. Izmailov, A.$\,$A. Yanbekov,
and A.$\,$A. Shayakhmetov, \emph{Ring-down gravitational waves and
lensing observables: How far can a wormhole mimic those of a black
hole?}, Phys. Rev. D \textbf{95}, 104011 (2017), \textcolor{purple}{\href{https://arxiv.org/abs/1611.03479}{arXiv:1611.03479 [gr-qc]}}

\bibitem{Bueno}P. Bueno, P.$\,$A. Cano, F. Goelen, T. Hertog, and
B. Vercnocke, \emph{Echoes of Kerr-like wormholes}, Phys. Rev. D \textbf{97},
024040 (2018), \textcolor{purple}{\href{https://arxiv.org/abs/1711.00391}{arXiv:1711.00391 [gr-qc]}}

\bibitem{Cramer}J.$\,$G. Cramer, R.$\,$L. Forward, M.$\,$S. Morris,
M. Visser, G. Benford, and G.$\,$A. Landis, \emph{Natural wormholes
as gravitational lenses}, Phys. Rev. D \textbf{51}, 3117 (1995), \textcolor{purple}{\href{https://arxiv.org/abs/astro-ph/9409051}{arXiv:astro-ph/9409051}}

\bibitem{Nedkova}P.$\,$G. Nedkova, V.$\,$K. Tinchev, and S.$\,$S.
Yazadjiev, \emph{Shadow of a rotating traversable wormhole}, Phys.
Rev. D \textbf{88}, 124019 (2013), \textcolor{purple}{\href{https://arxiv.org/abs/1307.7647}{arXiv:1307.7647 [gr-qc]}}

\bibitem{Harko}T. Harko, Z. Kovacs, and F.$\,$S.$\,$N. Lobo, \emph{Thin
accretion disks in stationary axisymmetric wormhole spacetimes}, Phys.
Rev. D \textbf{79}, 064001 (2009), \textcolor{purple}{\href{https://arxiv.org/abs/0901.3926}{arXiv:0901.3926 [gr-qc]}}

\bibitem{Deligianni}E. Deligianni, J. Kunz, P. Nedkova, S. Yazadjiev,
and R. Zheleva, \emph{Quasiperiodic oscillations around rotating traversable
wormholes}, Phys. Rev. D \textbf{104}, 024048 (2021), \textcolor{purple}{\href{https://arxiv.org/abs/2103.13504}{arXiv:2103.13504 [gr-qc]}}

\bibitem{Falco}V. De Falco, M. De Laurentis, and S. Capozziello,
\emph{Epicyclic frequencies in static and spherically symmetric wormhole
geometries}, Phys. Rev. D \textbf{104}, 024053 (2021), \textcolor{purple}{\href{https://arxiv.org/abs/2106.12564}{arXiv:2106.12564 [gr-qc]}}

\bibitem{Easson-2015}F. Duplessis and D.$\,$A. Easson, \emph{Exotica
ex nihilo: Traversable wormholes $\&$ non-singular black holes from
the vacuum of quadratic gravity}, Phys.$\,$Rev.$\,$D \textbf{92},
043516 (2015), \textcolor{purple}{\href{https://arxiv.org/abs/1506.00988}{arXiv:1506.00988 [gr-qc]}}

\bibitem{Easson-2017}J.$\,$B. Dent, D.$\,$A. Easson, T.$\,$W.
Kephart, and S.$\,$C. White, \emph{Stability Aspects of Wormholes
in $R^{2}$ Gravity}, Int.$\,$J.$\,$Mod.$\,$Phys.$\,$D \textbf{26},
1750117 (2017), \textcolor{purple}{\href{https://arxiv.org/abs/1608.00589}{arXiv:1608.00589 [gr-qc]}}

\bibitem{p4}K. Jusufi, A. \"Ovg\"un, A. Banerjee, \.I. Sakall{\i },
\emph{Gravitational lensing by wormholes supported by electromagnetic,
scalar, and quantum effects}, Eur. Phys. J. Plus \textbf{134}, 428
(2019), \textcolor{purple}{\href{https://arxiv.org/abs/1802.07680}{arXiv:1802.07680 [gr-qc]}}

\bibitem{p5}\.I. Sakall{\i }, A. \"Ovg\"un, \emph{Gravitinos
tunneling from traversable Lorentzian wormholes}, Astrophys Space
Sci \textbf{359}, 32 (2015), \textcolor{purple}{\href{https://arxiv.org/abs/1506.00599}{arXiv:1506.00599 [gr-qc]}}

\bibitem{p1}V. De Falco, E. Battista, S. Capozziello, and M. De Laurentis,
\emph{Reconstructing wormhole solutions in curvature based Extended
Theories of Gravity}, Eur. Phys. J. C \textbf{81}, 157 (2021), \textcolor{purple}{\href{https://arxiv.org/abs/2102.01123}{arXiv:2102.01123 [gr-qc]}}

\bibitem{p2}X.$\,$Y. Chew, B. Kleihaus, and J. Kunz, \emph{Spinning
Wormholes in Scalar-Tensor Theory}, Phys. Rev. D \textbf{97}, 064026
(2018), \textcolor{purple}{\href{https://arxiv.org/abs/1802.00365}{arXiv:1802.00365 [gr-qc]}}

\bibitem{p3}A. \"Ovg\"un, K. Jusufi, and \.I. Sakall{\i }, \emph{Exact
traversable wormhole solution in bumblebee gravity}, Phys. Rev. D
\textbf{99}, 024042 (2019), \textcolor{purple}{\href{https://arxiv.org/abs/1804.09911}{arXiv:1804.09911 [gr-qc]}}

\bibitem{p6}M.$\,$S. Churilova, R.$\,$A. Konoplya, Z. Stuchl\'{\i}k,
and A. Zhidenko, \emph{Wormholes without exotic matter: quasinormal
modes, echoes and shadows}, JCAP \textbf{10}, 010 (2021), \textcolor{purple}{\href{https://arxiv.org/abs/2107.05977}{arXiv:2107.05977 [gr-qc]}}

\bibitem{p7}G. Cl\'{e}ment and D. Gal'tsov, \emph{Rotating traversable
wormholes in Einstein--Maxwell theory}, Phys. Lett. B \textbf{838},
137677 (2023), \textcolor{purple}{\href{https://arxiv.org/abs/2210.08913}{arXiv:2210.08913 [gr-qc]}}

\bibitem{Agnese-1995}A.$\,$G. Agnese and M. La Camera, \emph{Wormholes
in the Brans-Dicke theory of gravitation}, Phys. Rev. D \textbf{51},
2011 (1995)

\bibitem{Agnese-2001}A.$\,$G. Agnese and M. La Camera, \emph{Schwarzschild
metrics, quasi-universes and wormholes}, in Sidharth, B.$\,$G., Altaisky,
M.$\,$V. (eds) \emph{Frontiers of Fundamental Physics 4}. Springer,
Boston, MA, \textcolor{purple}{\href{https://doi.org/10.1007/978-1-4615-1339-1\%5C\%5C_18}{https://doi.org/10.1007/978-1-4615-1339-1\_18}},
\textcolor{purple}{\href{https://arxiv.org/abs/astro-ph/0110373}{arXiv:astro-ph/0110373}}

\bibitem{Campanelli-1993}M. Campanelli and C. Lousto, \emph{Are Black
Holes in Brans-Dicke Theory precisely the same as in General Relativity?},
Int. J. Mod. Phys. D \textbf{2}, 451 (1993), \textcolor{purple}{\href{https://arxiv.org/abs/gr-qc/9301013}{arXiv:gr-qc/9301013}}

\bibitem{Vanzo-2012}L. Vanzo, S. Zerbini, and V. Faraoni, \emph{Campanelli-Lousto
and veiled spacetimes}, Phys. Rev. D \textbf{86}, 084031 (2012), \textcolor{purple}{\href{https://arxiv.org/abs/1208.2513}{arXiv:1208.2513 [gr-qc]}}

\bibitem{wec1}A.$\,$A. Sen and R.$\,$J. Scherrer, \emph{The weak
energy condition and the expansion history of the Universe}, Phys.
Lett. B \textbf{659}, 457 (2008), \textcolor{purple}{\href{https://arxiv.org/abs/astro-ph/0703416}{arXiv:0703416 [astro-ph]}}

\bibitem{wec2}J. Santos, J.$\,$S. Alcaniz, and M.$\,$J. Rebou\c{s}as,
\emph{Energy conditions and supernovae observations}, Phys. Rev. D
\textbf{74} (2006) 067301, \textcolor{purple}{\href{https://arxiv.org/abs/astro-ph/0608031}{arXiv:0608031 [astro-ph]}}

\bibitem{wec3}R.$\,$M. Wald, \emph{Asymptotic behavior of homogeneous
cosmological models in the presence of a positive cosmological constant},
Phys. Rev. D \textbf{28} (1983) 2118

\bibitem{wec4}G.$\,$J. Galloway, \emph{The lorentzian splitting
theorem without the completeness assumption}, Journal of differential
geometry \textbf{29}, 373 (1983)

\bibitem{wec5}C. Barcel\'o and M. Visser, \emph{Scalar fields, energy
conditions and traversable wormholes}, Class. Quantum Grav. \textbf{17},
3843 (2000), \textcolor{purple}{\href{https://arxiv.org/abs/gr-qc/0003025}{arXiv:0003025 [gr-qc]}}

\bibitem{wec6}L.$\,$H. Ford and T.$\,$A. Roman, \emph{Classical
scalar fields and the generalized second law}, Phys. Rev. D \textbf{64},
024023 (2001), \textcolor{purple}{\href{https://arxiv.org/abs/gr-qc/0009076}{arXiv:0009076 [gr-qc]}}

\bibitem{wec7}A. Borde and A. Vilenkin, \emph{Violation of the weak
energy condition in inflating spacetimes}, Phys. Rev. D \textbf{56},
717 (1997), \textcolor{purple}{\href{https://arxiv.org/abs/gr-qc/9702019}{arXiv:9702019 [gr-qc]}}

\bibitem{wec8}S. Kar, N. Dadhich, and M. Visser, \emph{Quantifying
energy condition violations in traversable wormholes}, Pramana - J.
Phys. \textbf{63}, 859 (2004), \textcolor{purple}{\href{https://arxiv.org/abs/gr-qc/0405103}{arXiv:0405103 [gr-qc]}}

\bibitem{wec9}E. Curiel, \emph{A Primer on Energy Conditions} in
D. Lehmkuhl, G. Schiemann, E. Scholz (eds), \textcolor{purple}{\href{https://link.springer.com/book/10.1007/978-1-4939-3210-8}{Towards a Theory of Spacetime Theories. Einstein Studies}},
vol. 13, Birkh\"auser, New York, NY (2017) 

\bibitem{Buchdahl-1962}H.$\,$A. Buchdahl, \emph{On the Gravitational
Field Equations Arising from the Square of the Gaussian Curvature},
Nuovo Cimento\textbf{ 23}, 141 (1962), \textcolor{purple}{\href{https://link.springer.com/article/10.1007/BF02733549}{https://link.springer.com/article/10.1007/BF02733549}}

\bibitem{Nelson-2010}W. Nelson, \textcolor{black}{\emph{Static solutions
for fourth order gravity}}\textcolor{black}{, Phys. Rev. D }\textbf{\textcolor{black}{82}}\textcolor{black}{,
104026 (2010), }\textcolor{purple}{\href{https://arxiv.org/abs/1010.3986}{arxiv:1010.3986 [gr-qc]}}

\bibitem{Stelle-2015}H. L\"u, A. Perkins, C.$\,$N. Pope, and K.$\,$S.
Stelle, \emph{Black holes in higher-derivative gravity,} Phys.$\,$Rev.$\,$Lett.
\textbf{114}, 171601 (2015), \textcolor{purple}{\href{https://arxiv.org/abs/1502.01028}{arxiv:1502.01028 [hep-th]}};
H. L\"u, A. Perkins, C.$\,$N. Pope, and K.$\,$S. Stelle, \emph{Spherically
symmetric solutions in higher-derivative gravity}, Phys.$\,$Rev.$\,$D
\textbf{92}, 124019 (2015), \textcolor{purple}{\href{https://arxiv.org/pdf/1508.00010.pdf}{arXiv:1508.00010 [hep-th]}}

\bibitem{Lust-2015-backholes}A. Kehagias, C. Kounnas, D. L\"ust,
and A. Riotto, \emph{Black hole solutions in $R^{2}$ gravity}, J.$\,$High
Energy Phys. \textbf{05}, 143 (2015), \textcolor{purple}{\href{https://arxiv.org/pdf/1502.04192.pdf}{arXiv:1502.04192 [hep-th]}}

\bibitem{Pravda-2017}V. Pravda, A. Pravdov\'a, J. Podolsk\'y, and
R. \v{S}varc, \emph{Exact solutions to quadratic gravity}, Phys.$\,$Rev.$\,$D
\textbf{95}, 084025 (2017), \textcolor{purple}{\href{https://arxiv.org/pdf/2012.08551.pdf}{arXiv:1606.02646 [gr-qc]};
}J. Podolsk\'y, R. \v{S}varc, V. Pravda, and A. Pravdov\'a, \emph{Explicit
black hole solutions in higher-derivative gravity}, Phys.$\,$Rev.$\,$D
\textbf{98}, 021502 (2018), \textcolor{purple}{\href{https://arxiv.org/abs/1806.08209}{arXiv:1806.08209 [gr-qc]}}

\bibitem{Gurses-2012}M. G\"urses, T.\c{C}. \c{S}i\c{s}man, and
B. Tekin, \emph{New exact solutions of quadratic curvature gravity},
Phys.$\,$Rev.$\,$D \textbf{86}, 024009 (2012); \textcolor{purple}{\href{https://arxiv.org/abs/1204.2215}{arXiv:1204.2215 [hep-th]}}

\bibitem{Alvarez-2018}E. Alvarez, J. Anero, S. Gonzalez-Martin, and
R. Santos-Garcia, \emph{Physical content of quadratic gravity,} Eur.$\,$Phys.$\,$J.$\,$C
\textbf{78}, 794 (2018), \textcolor{purple}{\href{https://arxiv.org/pdf/1802.05922.pdf}{arXiv:1802.05922 [hep-th]}}

\bibitem{Stelle-1977}K.$\,$S. Stelle, \emph{Renormalization of higher-derivative
quantum gravity,} Phys.$\,$Rev.$\,$D \textbf{16}, 953 (1977); K.$\,$S.
Stelle, \emph{Classical Gravity with Higher Derivatives}, Gen.$\,$Rel.$\,$Grav.
\textbf{9}, 353 (1978)

\bibitem{Frolov-2009}V.$\,$P. Frolov and I.$\,$L. Shapiro, \emph{Black
Holes in Higher Dimensional Gravity Theory with Quadratic in Curvature
corrections}, Phys.$\,$Rev.$\,$D \textbf{80}, 044034 (2009), \textcolor{purple}{\href{https://arxiv.org/abs/0907.1411}{arXiv:0907.1411 [gr-qc]}}

\bibitem{Murk-2022}S. Murk, \emph{Physical black holes in fourth-order
gravity}, Phys.$\,$Rev.$\,$D \textbf{105}, 044051 (2022), \textcolor{purple}{\href{https://arxiv.org/abs/2110.14973}{arXiv:2110.14973 [gr-qc]}}

\bibitem{Clifton-2006}T. Clifton, \emph{Spherically Symmetric Solutions
to Fourth-Order Theories of Gravity}, Class.$\,$Quant.$\,$Grav.
\textbf{23}, 7445 (2006), \textcolor{purple}{\href{https://arxiv.org/pdf/gr-qc/0607096.pdf}{arXiv:gr-qc/0607096}}

\bibitem{Rinaldi-2018}M. Rinaldi, \emph{On the equivalence of Jordan
and Einstein frames in scale-invariant gravity}, Eur.$\,$Phys.$\,$J.$\,$Plus
\textbf{133}, 408 (2018), \textcolor{purple}{\href{https://arxiv.org/pdf/1808.08154.pdf}{arXiv:1808.08154 [gr-qc]}}

\bibitem{Donoghue-2018}J.$\,$F. Donoghue and G. Menezes, \emph{Gauge
assisted quadratic gravity: A framework for UV complete quantum gravity},
Phys.$\,$Rev.$\,$D \textbf{97}, 126005 (2018), \textcolor{purple}{\href{https://arxiv.org/pdf/1804.04980.pdf}{arXiv:1804.04980 [hep-th]}}

\bibitem{Ferreira-2019}P.$\,$G. Ferreira and O.$\,$J. Tattersall,
\emph{Scale invariant gravity and black hole ringdown}, Phys.$\,$Rev.$\,$D
\textbf{101}, 024011 (2020), \textcolor{purple}{\href{https://arxiv.org/pdf/1910.04480.pdf}{arXiv:1910.04480 [gr-qc]}}

\bibitem{Nguyen-2022-Buchdahl}H.$\,$K. Nguyen, \emph{Beyond Schwarzschild-de
Sitter spacetimes: I. A new exhaustive class of metrics inspired by
Buchdahl for pure $R^{2}$ gravity in a compact form}, Phys.$\,$Rev.$\,$D
\textbf{106}, 104004 (2022),\textcolor{purple}{{} \href{https://arxiv.org/abs/2211.01769}{arXiv:2211.01769 [gr-qc]}}

\bibitem{Nguyen-2023-Lambda0}H.$\,$K. Nguyen, \emph{Beyond Schwarzschild-de
Sitter spacetimes: II. An exact non-Schwarzschild metric in pure $R^{2}$
gravity and new anomalous properties of $R^{2}$ spacetimes}, Phys.$\,$Rev.$\,$D
\textbf{107}, 104008 (2023), \textcolor{purple}{\href{https://arxiv.org/abs/2211.03542}{arXiv:2211.03542 [gr-qc]}}

\bibitem{Nguyen-2023-Extension}H.$\,$K. Nguyen, \emph{Beyond Schwarzschild-de
Sitter spacetimes: III. A perturbative vacuum with non-constant scalar
curvature in $R+R^{2}$ gravity}, Phys.$\,$Rev.$\,$D \textbf{107},
104009 (2023),\textcolor{purple}{{} \href{https://arxiv.org/abs/2211.07380}{arXiv:2211.07380 [gr-qc]}}

\bibitem{Nguyen-2023-Nontrivial}H.$\,$K. Nguyen, \emph{Non-triviality
of asymptotically flat Buchdahl-inspired metrics in pure $\mathcal{R}^{2}$
gravity},\textcolor{purple}{{} \href{https://arxiv.org/abs/2305.12037}{arXiv:2305.12037 [gr-qc]}}

\bibitem{2023-axisym}M. Azreg-A\"inou and H.$\,$K. Nguyen, \emph{A
stationary axisymmetric vacuum solution for pure $R^{2}$ gravity},
\textcolor{purple}{\href{https://arxiv.org/abs/2304.08456}{arXiv:2304.08456 [gr-qc]}}

\bibitem{2023-WH}H.$\,$K. Nguyen and M. Azreg-A\"inou, \emph{Traversable
Morris-Thorne-Buchdahl wormholes in quadratic gravity}, \textcolor{purple}{\href{https://arxiv.org/abs/2305.04321\%5C\%20}{arXiv:2305.04321 [gr-qc]}}

\bibitem{Edery-2014}A. Edery and Y. Nakayama, \emph{Restricted Weyl
invariance in four-dimensional curved spacetime}, Phys.$\,$Rev.$\,$D
\textbf{90}, 043007 (2014), \textcolor{purple}{\href{https://arxiv.org/abs/1406.0060}{arXiv:1406.0060 [hep-th]}}

\bibitem{Woodard}R.$\,$P. Woodard, \emph{Avoiding Dark Energy with
$1/R$ Modifications of Gravity}, Lect.$\,$Notes Phys. \textbf{720},
403 (2007), \textcolor{purple}{\href{https://arxiv.org/pdf/astro-ph/0601672.pdf}{arXiv:astro-ph/0601672};
}R.$\,$P. Woodard, \emph{Ostrogradsky\textquoteright s theorem on
Hamiltonian instability}, Scholarpedia \textbf{10}, 32243 (2015),\textcolor{purple}{{}
\href{https://arxiv.org/pdf/1506.02210.pdf}{arXiv:1506.02210 [hep-th]}}

\bibitem{AlvarezGaume-2015}L. Alvarez-Gaume, A. Kehagias, C. Kounnas,
D. L\"ust, and A. Riotto, \emph{Aspects of Quadratic Gravity}, Fortsch.$\,$Phys.
\textbf{64}, 176 (2016), \textcolor{purple}{\href{https://arxiv.org/pdf/1505.07657.pdf}{arXiv:1505.07657 [hep-th]}}

\bibitem{Nguyen-2023-Newtonian}H\@.$\,$K. Nguyen, \emph{Emerging
Newtonian potential in pure $\mathcal{R}^{2}$ gravity on a de Sitter
background},\emph{ }\textcolor{purple}{\href{https://arxiv.org/abs/2306.03790}{arXiv:2306.03790 [gr-qc]}}

\bibitem{2023-WEC}H.$\,$K. Nguyen and M. Azreg-A\"inou, \emph{New
insights into Weak Energy Condition and wormholes in Brans-Dicke gravity}\textcolor{purple}{,
\href{https://arxiv.org/abs/2305.15450}{arXiv:2305.15450 [gr-qc]}}

\bibitem{Koutou}E-A. Kontou and K. Sanders, \emph{Energy conditions
in general relativity and quantum field theory}, Class. Quantum. Grav.
37 (2020), 193001, \textcolor{purple}{\href{https://arxiv.org/abs/2003.01815}{arXiv:2003.01815 [gr-qc]}}

\bibitem{distinguish}N. Tsukamoto, T. Harada, and K. Yajima, \emph{Can
we distinguish between black holes and wormholes by their Einstein
ring systems?}, Phys.$\,$Rev.$\,$D \textbf{86}, 104062 (2012), \textcolor{purple}{\href{https://arxiv.org/abs/1207.0047}{arXiv:1207.0047 [gr-qc]}};
C. Bambi and D. Stojkovic, \emph{Astrophysical Wormholes}, Universe
\textbf{7}, 136 (2021), \textcolor{purple}{\href{https://arxiv.org/abs/2105.00881}{arXiv:2105.00881 [gr-qc]}}

\bibitem{future}K. Jusufi, M. Azreg-A\"inou, and M. Jamil, \emph{Orbital
motions in the special Buchdahl-inspired metric} (in preparation)

\bibitem{st1}M. Azreg-A\"inou, \emph{Instability of two-dimensional
heterotic stringy black holes}, Class. Quantum. Grav. 16 (1999) 245,
\textcolor{purple}{\href{https://arxiv.org/abs/gr-qc/9902005}{arXiv:9902005 [gr-qc]}}

\bibitem{st2}M. Azreg-A\"inou, G. Cl\'ement, C.$\,$P. Constantinidis,
J.$\,$C. Fabris, \emph{Electrostatic solutions in Kaluza-Klein theory:
geometry and stability}, Grav. Cosmol. 6 (2000) 207, \textcolor{purple}{\href{https://arxiv.org/abs/gr-qc/9911107}{arXiv:9911107 [gr-qc]}}

\bibitem{s1}K. Bronnikov, S. Bolokhov, A. Makhmudov, and M. Skvortsova\emph{,
``Trapped ghost'' wormholes and regular black holes. The stability
problem}, \textcolor{purple}{\href{https://indico.cern.ch/event/597202/contributions/2714705/attachments/1531786/2397765/Bronnikov-Yerevan.pdf}{Modern Physics of Compact Stars and Relativistic Gravity, 2017, Yerevan}}

\bibitem{s2}F.$\,$S.$\,$N. Lobo, \emph{Stability of phantom wormholes},
Phys. Rev. D 71 (2005) 124022, \textcolor{purple}{\href{https://arxiv.org/abs/gr-qc/0506001}{arXiv:0506001 [gr-qc]}}

\bibitem{s3}D.$\,$I. Novikov, A.$\,$G. Doroshkevich, I.$\,$D.
Novikov, and A.$\,$A. Shatskii, \emph{Semi-permeable wormholes and
the stability of static wormholes}, Astron. Rep. 53, (2009), 1079
\textcolor{purple}{\href{https://arxiv.org/abs/0911.4456}{arXiv:0911.4456 [gr-qc]}}

\bibitem{s4}K.$\,$A. Bronnikov and S.$\,$G. Rubin, \textcolor{purple}{\href{https://doi.org/10.1142/12186}{Black holes, cosmology and extra dimensions}},
2nd edition, World Scientific, Singapore, 2022
\end{thebibliography}
\end{document}